\documentclass[12pt]{elsarticle}
\usepackage{slashed}
\usepackage{epsfig,graphics}
\usepackage{hyperref}
\usepackage{bbm}
\usepackage{mathtools}
\usepackage{amssymb}
\usepackage{mathrsfs}
\usepackage{amsmath}
\usepackage{natbib}

\newlength{\llslash}

\begin{document}

\begin{frontmatter}
\title{On pion and kaon decay constants \\ and chiral SU(3) extrapolations}
\author[GSI]{Xiao-Yu Guo}
\author[GSI,TUD]{M.F.M. Lutz}
\address[GSI]{GSI Helmholtzzentrum f\"ur Schwerionenforschung GmbH,\\
Planck Str. 1, 64291 Darmstadt, Germany}
\address[TUD]{Technische Universit\"at Darmstadt, D-64289 Darmstadt, Germany}
\begin{abstract}
We consider the dependence of the pion and kaon decay constants on the up, down and strange quark masses in QCD with strict isospin symmetry. The role of dynamical vector meson degrees of freedom is scrutinized in 
terms of an effective chiral Lagrangian for vector mesons. Applying a set of low-energy parameters as determined previously from 
QCD lattice data on the masses of the light vector mesons from PACS-CS, QCDSF-UKQCD and HSC we compute its implications on the pion and kaon decay constants for QCD lattice ensembles of HPQCD, CLS and ETMC. It is shown that with Gasser-Leutwyler  $L_4$ and $L_5$ parameters fixed to the empirical decay constants an accurate reproduction of their values at unphysical quark masses as computed by HPQCD, CLS and ETMC is achieved. Results for the masses of the light vector meson, the $\omega -\phi$ mixing angles and  the quark  mass ratios for the ensembles used by  HPQCD, CLS and ETMC  are discussed. 
\end{abstract}
\end{frontmatter}

\section{Introduction}

So far any chiral extrapolation attempt for the decay constants of the Goldstone bosons of QCD that is based on the chiral Lagrangian formulated for three light flavours appeared futile because of the rather large strange quark mass.
By today most lattice collaborations abandoned the use of three flavour extrapolations and consider  chiral extrapolations significant only in the small up and down quark masses of QCD \cite{Aoki:2016frl}. In turn the flavour SU(3) limit value, $f$, of the pion and kaon decay constants is poorly known at present. Given the fact that $\Lambda_\chi =4\,\pi \,f$  sets the chiral symmetry breaking scale of QCD with three light flavours this is quite unfortunate. The parameter $f$ is of fundamental importance in hadronic interactions since it drives any application of the three-flavour chiral Lagrangian of QCD \cite{Ecker:1988te,Jenkins:1995vb,Klingl:1996by,Birse:1996hd,Bijnens:1997ni,Bijnens:1997rv,Cirigliano:2003yq,Rosell:2004mn,Bruns:2013tja}. The root of  this unpleasant situation lies in the rather poor convergence properties of a strict $\chi$PT  expansion based on the three-flavour chiral Lagrangian \cite{DescotesGenon:2003cg,Soto:2011ap,Ecker:2013pba,Guo:2014yva,Bijnens:2014lea,Terschlusen:2016cfw}.

In this work we present a remedy of this issue by two unconventional ingrediences. First we recast one-loop expressions derived from a chiral Lagrangian in terms of physical meson masses, while keeping the renormalization scale invariance of the 
expressions \cite{Lutz:2018cqo,Guo:2018kno,Bavontaweepanya:2018yds}. Second, we consider a chiral Lagrangian with explicit vector meson degrees of freedom \cite{Lutz:2008km,Terschlusen:2012xw,Bavontaweepanya:2018yds}. Its low-energy constants have been determined recently at the one-loop level in \cite{Guo:2018b} from  
lattice data on meson masses from PACS-CS, QCDSF-UKQCD and HSC \cite{PACS-CS,QCDSF-UK,HSC,Dudek:2013yja}. In particular a rather small value for $f = (70.5\pm 3.0)$ MeV was obtained compatible with the range suggested by the two-loop estimates of $\chi$PT in \cite{Bijnens:2014lea}. This study 
left undetermined the two low-energy parameters $L_4$ and $L_5$ only, which enter the computation of the pion and kaon decay constants. It is the purpose of this work to present results for the decay constants on the ensembles used by HPQCD, CLS and ETMC \cite{Dowdall:2013rya,Bruno:2016plf,Ottnad:2017bjt}.

\section{Chiral dynamics for the pion and kaon decay constants}

Our analysis on the pion and kaon decay constants is based on the chiral Lagrangian with dynamical vector meson degrees of freedom as further developed recently in \cite{Bavontaweepanya:2018yds,Guo:2018b}. For the specifics of the Lagrangian and explicit expressions valid at the one-loop level properly derived in a finite volume  we refer to our previous works \cite{Bavontaweepanya:2018yds,Guo:2018b}. Note that as in \cite{Bavontaweepanya:2018yds} we do not yet consider the explicit effects of the $ \eta'$ field \cite{Terschlusen:2012xw}.

\begin{figure}[t]
\centering
\includegraphics[width=0.95\textwidth]{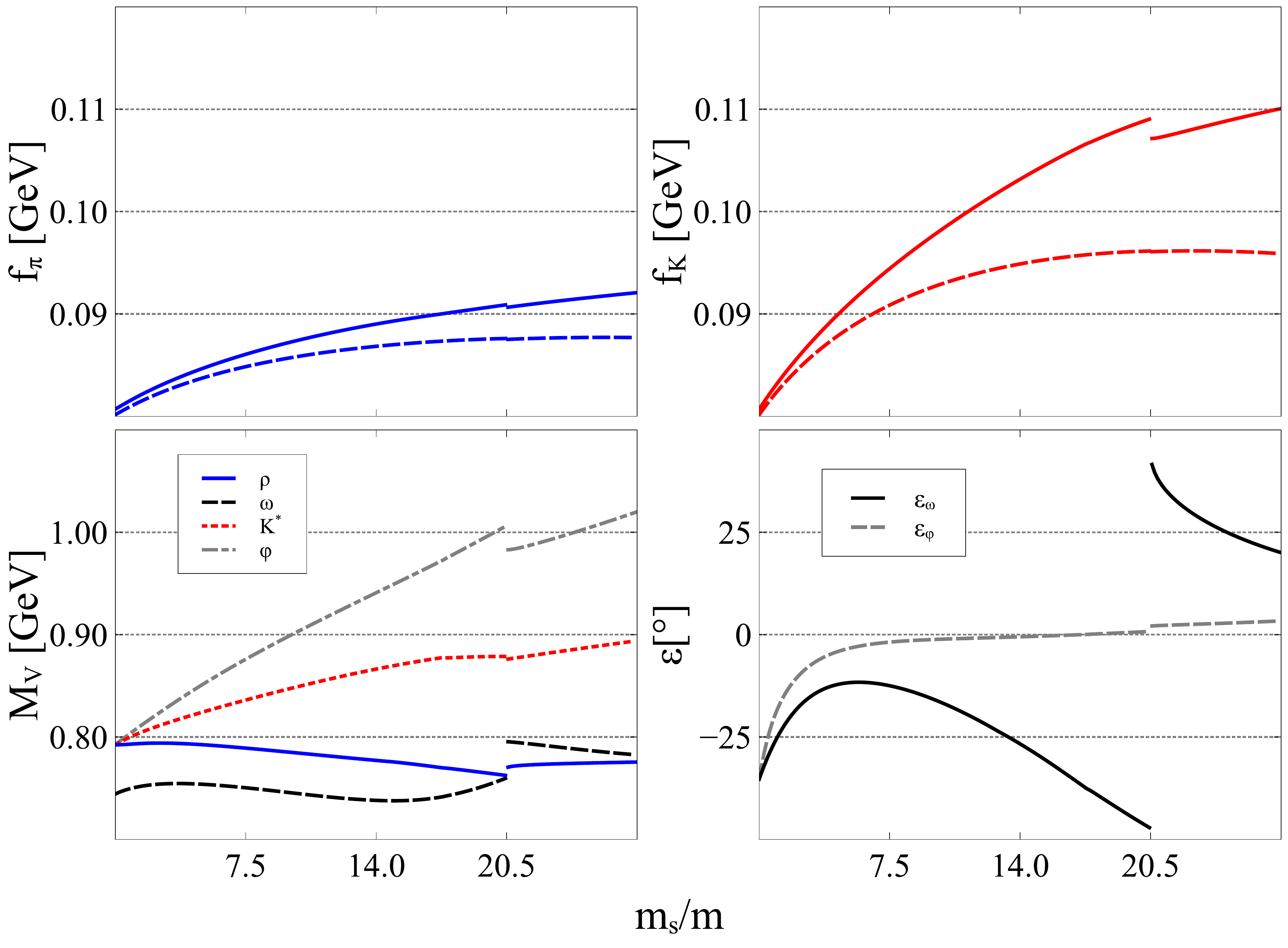} 
\caption{Our results from Fit 2 of \cite{Guo:2018b} for the pion and kaon decay constants in the infinite volume limit as a function of the quark mass ratio $m_s/m$ at physical value for $m$. While the solid lines are from (1), the dashed ones from (2) where the contributions of vector meson loops are ignored. Further results are shown for the masses and mixing angles of the vector mesons. }
\label{fig:0}
\end{figure} 
 
Here we recall only the expressions for the decay constants. The one-loop contributions from the Goldstone bosons and the light vector mesons in (\ref{def-fP}) drive the decay constants $f_P$ away from their chiral limit value $f$.  Altogether, for the decay constants we use the simple result
\begin{eqnarray}
	&& f_P = f^{\chi -\rm PT}_P + \Big(\sqrt{Z_P^{\rm bubble}} -1 \Big)f -  \frac{f}{\sqrt{Z_P^{\rm bubble}}} \,\frac{\Pi_P^{\rm bubble}}{m_P^2} \,,
	\label{f_Def_mu-Indep}
\nonumber\\
	&& Z_P^{\rm bubble} = 1 + \frac{\partial}{\partial\, m_P^2}\,\Pi_P^{\rm bubble}(m_P^2) \,,
	\label{def-fP}
\end{eqnarray}
where we split the loop contributions into a conventional part  $ f^{\chi -\rm PT}_P $ and terms that reflect the presence of dynamical vector meson degrees of freedom. The contribution from the latter was derived previously in \cite{Bavontaweepanya:2018yds} at $Z_P^{\rm bubble} =1$. Here we consider the effect from the wave function renormalization in addition.
The conventional part 
\begin{eqnarray}
&&f\, f^{\chi -\rm PT}_\pi = f^2 -\bar I_\pi   - \frac{1}{2}\,\bar I_K + 4\,m_\pi^2\, L_5 + 4\,(2\,m_K^2+ m_\pi^2 )\,L_4\,,
\nonumber\\
&& f\,f^{\chi -\rm PT}_K = f^2- \frac{3}{8}\,\bar I_\pi   - \frac{3}{4}\,\bar I_K - \frac{3}{8}\,\bar I_\eta  +  4\,m_K^2\, L_5 +  6\,(m_\eta^2+ m_\pi^2 )\,L_4 \,,
\nonumber\\
&& \bar I_P = \frac{m_P^2}{(4\pi)^2}\,\log \frac{m_P^2}{\mu^2} + {\rm finite\, volume \,terms}\,,
\label{def-fP-chiPT}
\end{eqnarray}
involves the 
low-energy constants $L_4$ and $L_5$ of Gasser and Leutwyler and the tadpole integrals $\bar I_P$  properly evaluated in a finite volume. The latter depends on the meson mass $m_P$, the renormalization scale $\mu$ and the lattice volume only (see e.g \cite{Lutz:2014oxa}). Our expressions differ from the traditional results to the extent that our form does not involve any explicit dependence on the quark masses. Still the expressions do not depend on the renormalization scale $\mu$, strictly. This reflects our strategy that hadron masses inside loop expressions should take their on-shell values. We assure that upon a  further chiral expansion of $f^{\chi -\rm PT}_P$ the traditional form is recovered identically. The contributions of the vector mesons are implied by a bubble loop contributions to the polarization function $\Pi_P(s = m_P^2)$ as was derived in \cite{Bavontaweepanya:2018yds,Guo:2018b}. The latter involves the masses of the Goldstone bosons, $m_P$, and the vector mesons, $M_V$, in their isospin limit.

In Fig. \ref{fig:0} the effect of dynamical vector mesons is illustrated as it comes in the infinite volume limit. In the upper panel the pion and kaon decay constants are shown as a function of the quark mass ratio $r=m_s/m$ at physical value of $m=m_u=m_d$. While at the flavour limit point with $r=1$, the results from (1) and (2) almost agree, at the physical ratio with $r \simeq 26.5$ we find a significant effect from the vector meson loop contributions.
We note a discontinuous dependence on the quark mass ratio. A small but significant jump in the kaon decay constant is visible at $r \simeq 20.5$. The main driving source of this effect is traced to a striking discontinuous behaviour of the mass dependent $\omega-\phi$ mixing angle. Of physical relevance is the mixing angle evaluated at either the $\omega$ or $\phi$ on-shell mass only, which we denote by $\epsilon_\omega$ and $\epsilon_\phi $ respectively.
The latter are shown  in the lower panels of Fig. \ref{fig:0} which also presents our results for the vector meson masses. Here the $\omega$ and $\phi$ meson masses exhibit a visible jump at the same critical value of the quark-mass ratio. It would be important to scrutinize our prediction by suitable QCD lattice simulations for the quark-mass depedence of the $\omega-\phi$ mixing angles. 
The possibility of such a discontinuous behaviour opens once on-shell masses inside loop expressions are used \cite{Semke:2006hd}. This implies that a set of non-linear and coupled equations have to be considered. We emphasize that such a phase transition cannot be ruled out from first principals in QCD. Similar transitions, but in different systems, were observed in previous works \cite{Semke:2006hd,Guo:2011gc}.

In the determination of our low-energy constants \cite{Guo:2018b} from QCD lattice data  we considered finite-box energy levels from PACS-CS, QCDSF-UKQCD and HSC \cite{PACS-CS,QCDSF-UK,HSC,Dudek:2013yja}. 
For any lattice ensemble of a given finite-box size we took the pion and kaon masses as input parameters. The set of nine coupled and non-linear mass equations is solved in terms of the two quark masses, $B_0\, m$, $B_0\, m_s$,  the remaining 5 meson masses, $m_\eta, M_\rho, M_\omega, M_{K^*}, M_\phi$ and the $\omega-\phi$ mixing angles $\epsilon_\omega$ and $\epsilon_\phi $.  We apply the evolutionary algorithm of GENEVA 1.9.0-GSI \cite{Geneva} with runs of a population size 1500 on 300 parallel CPU cores. 
The two parameters $L_4$ and $L_5$ can be dialed always as to recover the PDG values of $f_\pi = (92.1 \pm 1.2)$ MeV and $f_K =( 110.0\pm 0.3)$ MeV. The challenge is to describe then the decay constants at unphysical quark masses as provided by QCD lattice computations.  We emphasize that the presence of the vector meson contributions as given here and \cite{Guo:2018b} does not renormalize either $L_4$ or $L_5$. Given our renormalization scheme the vector meson contributions are at least of quadratic order in the quark masses. The decay constants depend on $f, L_4, L_5$ and the meson masses only. No further explicit unknown parameter dependence is encountered in our one-loop framework.

Following the strategy of our previous works  \cite{Lutz:2018cqo,Guo:2018kno,Guo:2018b} we use the empirical values of the meson masses and together with the pion and kaon decay constants from the 
Particle Data Group \cite{PDG} as additional constraints in our fit scenarios. The main target of our studies is to derive its low-energy representation in terms of hadronic degrees of freedom. 
In this case it is of advantage to perform a non-standard scale setting in terms of a larger set of observable quantities. The lattice scale of all ensembles at a given $\beta $ value is considered as a free parameter to be determined from the lattice data set together with the chosen set of quantities from the PDG.

\begin{table}[t]
\setlength{\tabcolsep}{6.5mm}
\renewcommand{\arraystretch}{1.2}
\begin{center}
\begin{tabular}{l|rrr|r} 
                             & Fit 1      & Fit 2    & Fit 3  & Literature     \\ \hline 
$f$ [MeV]                    & 73.57      & 70.72    &   67.51        &   64 - 71 \cite{Bijnens:2014lea} \\
$L_4 \times 10^3$            &  -0.0358   & 0.0485   &  0.1338      &  0.3 - 0.76  \cite{Bijnens:2014lea}   \\
$L_5 \times 10^3$            &  -0.0117   & 0.0192   &  0.0364      &  0.50 - 1.01  \cite{Bijnens:2014lea} \\ 
\end{tabular}
\caption{Selection of low-energy parameters from the three fit scenarios of \cite{Guo:2018b}. $L_4$ and $L_5$ are given at the renormalization scale $\mu = 0.77$ GeV. }
\end{center}
\label{tab:parameter:A}
\end{table}

In Tab. 1 of \cite{Guo:2018b} three sets of LEC are collected. While Fit 1 is based on the meson masses only, the other two scenarios considered the pion and kaon decay constants as measured by HPQCD and CLS  on their lattice ensembles  \cite{Dowdall:2013rya,Bruno:2016plf}. Our Fit 3 considers in addition the pion decay constants from ETMC \cite{Ottnad:2017bjt}. Here we do not take into account their kaon decay constants, since they are affected by a wave function factor that is subject to significant uncertainties. We will return to this issue below.

\begin{table}[t]
\setlength{\tabcolsep}{6.8mm}
\renewcommand{\arraystretch}{1.3}
\begin{center}
\begin{tabular}{l|rr||l} 
                                      & Fit 2      &  Fit 3   &   Lattice  \\ \hline 
$a^{\beta = 5.8}_{\rm HPQCD}$ [fm]    &  0.1535    &  0.1524  &   0.1509 - 0.1543 \\
$\chi^2/N$                            &  1.11      &  1.50    &   \cite{Dowdall:2013rya} \\ \hline
$a^{\beta = 6.0}_{\rm HPQCD}$ [fm]    &  0.1230    &  0.1222  &   0.1212 - 0.1241 \\
$\chi^2/N$                            &  0.93      &  1.39    &    \cite{Dowdall:2013rya} \\ \hline
$a^{\beta = 6.3}_{\rm HPQCD}$ [fm]    &  0.0890    &  0.0887  &   0.0879 - 0.0907 \\ 
$\chi^2/N$                            &  1.18      &  1.37    &    \cite{Dowdall:2013rya} \\ \hline \\ \hline
$a^{\beta = 3.40}_{\rm CLS}$ [fm]     &  0.0786    &  0.0778  &   I : 0.08636(98)(40)\\
$\chi^2/N$                            &  0.33      &  0.51    &   II: 0.0790(11)\hfill \cite{Bruno:2016plf} \\ \hline
$a^{\beta = 3.46}_{\rm CLS}$ [fm]     &  0.0715    &  0.0706  &   I : 0.07634(92)(31)   \\
$\chi^2/N$                            &  0.20      &  0.15    &   II: 0.071(2) \hfill \cite{Bruno:2016plf} \\ \hline
$a^{\beta = 3.55}_{\rm CLS}$ [fm]     & 0.0603     & 0.0598   &   I : 0.06426(74)(17)  \\
$\chi^2/N$                            & 1.00       & 1.30     &   II: 0.0613(9) \hfill \cite{Bruno:2016plf} \\ \hline
$a^{\beta = 3.70}_{\rm CLS}$ [fm]     & 0.0475     & 0.0471   &   I : 0.04981(56)(10) \\ 
$\chi^2/N$                            & 0.23       & 0.27     &   II: 0.0481(8)\hfill \cite{Bruno:2016plf} \\ \hline \\ \hline
$a^{\beta = 1.95}_{\rm ETM}$ [fm]     &  0.0830    &  0.0830  &   0.0815(30)  \\
$\chi^2/N$                            &  20.86     &  7.06    &   \cite{Ottnad:2017bjt} \\ \hline
$a^{\beta = 2.10}_{\rm ETM}$ [fm]     &  0.0610    &  0.0610  &   0.0619(18) \\
$\chi^2/N$                            &  1.79      &  0.98    &  \cite{Ottnad:2017bjt}  \\ \hline
\end{tabular}
\caption{Partial $\chi^2/N$ values and lattice scales for the various ensembles as implied by the fit scenario 2 and 3.  }
\end{center}
\label{tab:parameter:B}
\end{table}

In Tab. 1 the result of our three scenarios  for $f$, $L_4$ and $L_5$ are displayed. 
A rough estimate of the uncertainties in the LEC is suggested by the spread of the latter in  the three scenarios. Four significant digits are shown in order to permit a numerical reproduction. The values in $f$ and $L_4, L_5$ show  moderate variations. A rather small value for $L_4$ is obtained always as was expected by its suppression in the large-$N_c$ limit of QCD. We observe that the strong anti-correlation of the parameters $f$ and $L_4$ 
in the decay constants as emphasized in  \cite{DescotesGenon:2003cg,Ecker:2013pba}, is lifted significantly in our approach since it considers QCD lattice data at unphysical quark masses. 
We find most remarkable the small values for $L_5$ in the three scenarios, which are in striking conflict with the range provided by the conventional approach based on the chiral SU(3) Lagrangian at the two-loop level \cite{Bijnens:2014lea}. The 
consideration of dynamical vector meson degrees of freedom causes a significant change in $L_5$, driving it to a value that is almost compatible with zero at the given renormalization scale. This was anticipated already in our previous work \cite{Bavontaweepanya:2018yds} and should be scrutinzed in further dedicated lattice studies.

Since a large set of lattice data is fitted the propagated statistical error on any of the low-energy constants is very small and insignificant. Any uncertainty in the latter stems from systematical deficiencies underlying our approach, like the neglect of discretization effects, or the impact of two-loop diagrams in our scheme. Such a systematic study is much beyond the scope of the present work. Note that the small uncertainties in the empirical values for the vector meson masses or the decay constants do not propagate to any  significant uncertainty in our scheme.

\begin{figure}[t]
\centering
\includegraphics[width=0.95\textwidth]{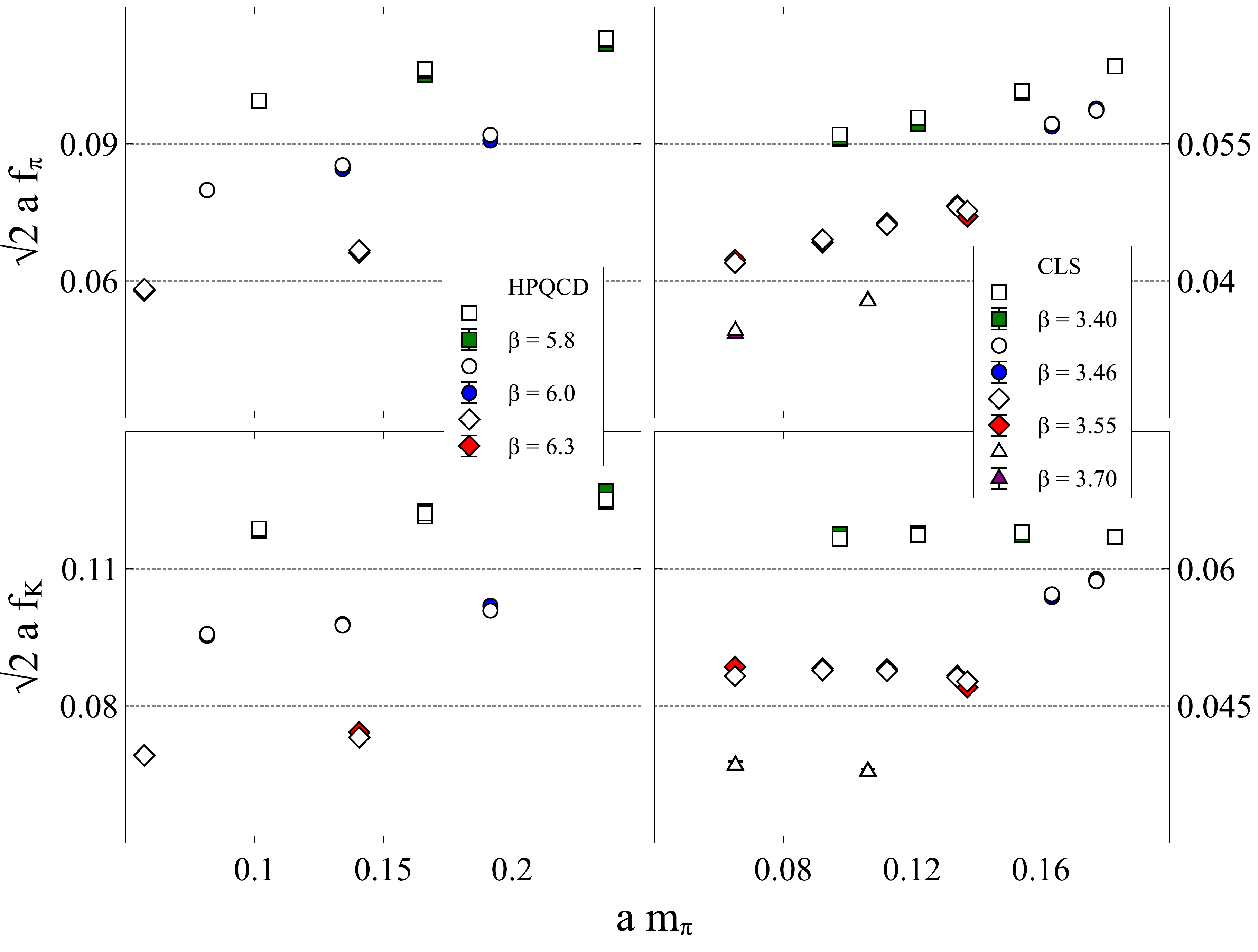}
\caption{Our results from Fit 2 for the pion and kaon decay constants in lattice units, where the corresponding lattice scales are collected in Tab. \ref{tab:parameter:B}. The lattice results are given by green, blue, and  red  filled symbols, where statistical errors are shown only. They are compared to the chiral extrapolation results in open symbols, which are always displayed on top of the lattice symbols.   }
\label{fig:1}
\end{figure}

At this stage we assume universal systematical errors for the vector meson masses and the decay constants. Our fits are based on an asymmetric error in the vector meson masses of $^{+10}_{-20}$ MeV together with a symmetric error of $\pm 1.25$ MeV for the decay constants $\sqrt{2}\, f_P$. This implies the $\chi^2/N$ values collected in Tab. \ref{tab:parameter:B} and also in Tab. 1 of \cite{Guo:2018b}, where $N$ is always the number of fitted lattice data points. 
The lattice data sets from HPQCD and CLS on the pion and kaon decay constants
are well reproduced in Fit 2 and Fit 3. 
Our results from Fit 2 are visualized by Fig. \ref{fig:1}, in which we show the decay constants in units of the lattice scale for all considered ensembles of HPQCD and CLS. 
With typical values $\chi^2/N \sim 1$ the decay constants are recovered with an  uncertainty of about $0.9$ MeV. 

\begin{figure}[t]
\centering
\includegraphics[width=0.99\textwidth]{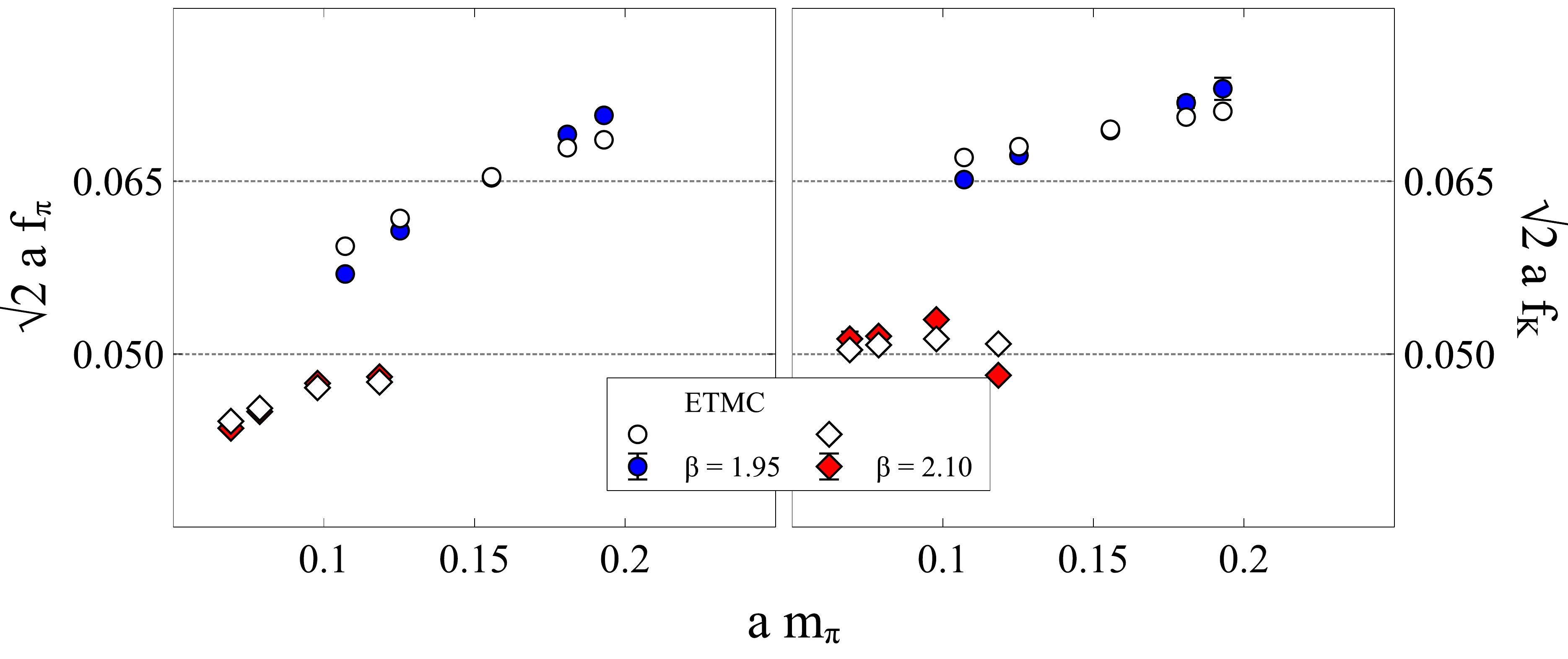}
\caption{Our results from Fit 3 for the pion and kaon decay constants in lattice units. The lattice results are given by blue and  red  filled symbols, where statistical errors are shown only. They are compared to the chiral extrapolation results in open symbols, which are always displayed on top of the lattice symbols. We use $Z=0.6884$ and $Z =0.7428$ for the $\beta=1.95$ and $\beta = 2.10$ ensembles respectively. Only statistical error bars are shown. }
\label{fig:2}
\end{figure}

In Tab. \ref{tab:parameter:B} we provide also our results for the lattice scales  of HPQCD and CLS at different $\beta $ values as they are a consequence of our fit strategy. 
While our scale setting for the three $\beta$ values considered by HPQCD appears compatible with the analysis in \cite{Dowdall:2013rya}, we observe a disagreement with the preferred scale setting for the four $\beta$ values of CLS in \cite{Bruno:2016plf}. The authors report on two distinct methods. Their method I, which has small statistical errors only, is in conflict to our results. However, their second method, which comes with somewhat larger statistical errors, predicts lattice scales that are quite compatible with our values. Both values are recalled in the last column of  Tab.  \ref{tab:parameter:B}. While two different scale setting methods need not lead to identical lattice scales, the size of discretization effects in the observable quantities may be distinct in the two methods. Our conclusion on the CLS ensembles would be that their second method, may have larger statistical errors, however, it comes with smaller 
discretization errors, and therefore is more convenient to use.

\begin{figure}[t]
\centering
\includegraphics[width=0.99\textwidth]{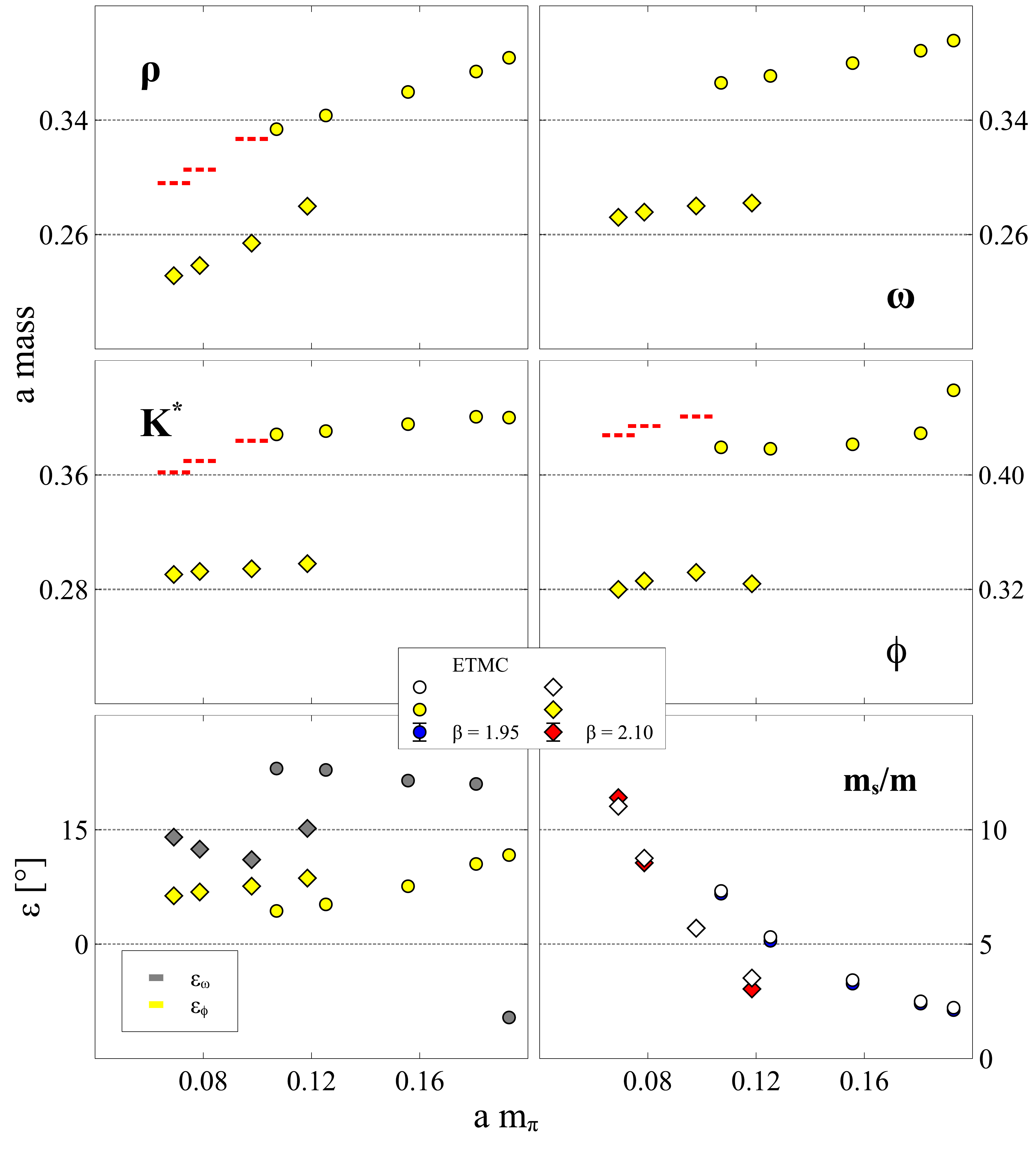}
\caption{Our results from Fit 3 for the vector meson masses, $\omega-\phi$ mixing angles and quark mass ratios on the ETM ensembles. The lattice results are given 
by blue and red filled symbols. They are compared to the chiral extrapolation results in open symbols, which are always displayed on top of the lattice symbols. We use yellow or grey colour filled symbols for the cases where there is no corresponding lattice point available yet. }
\label{fig:3}
\end{figure}

\begin{figure}[t]
\centering
\includegraphics[width=0.99\textwidth]{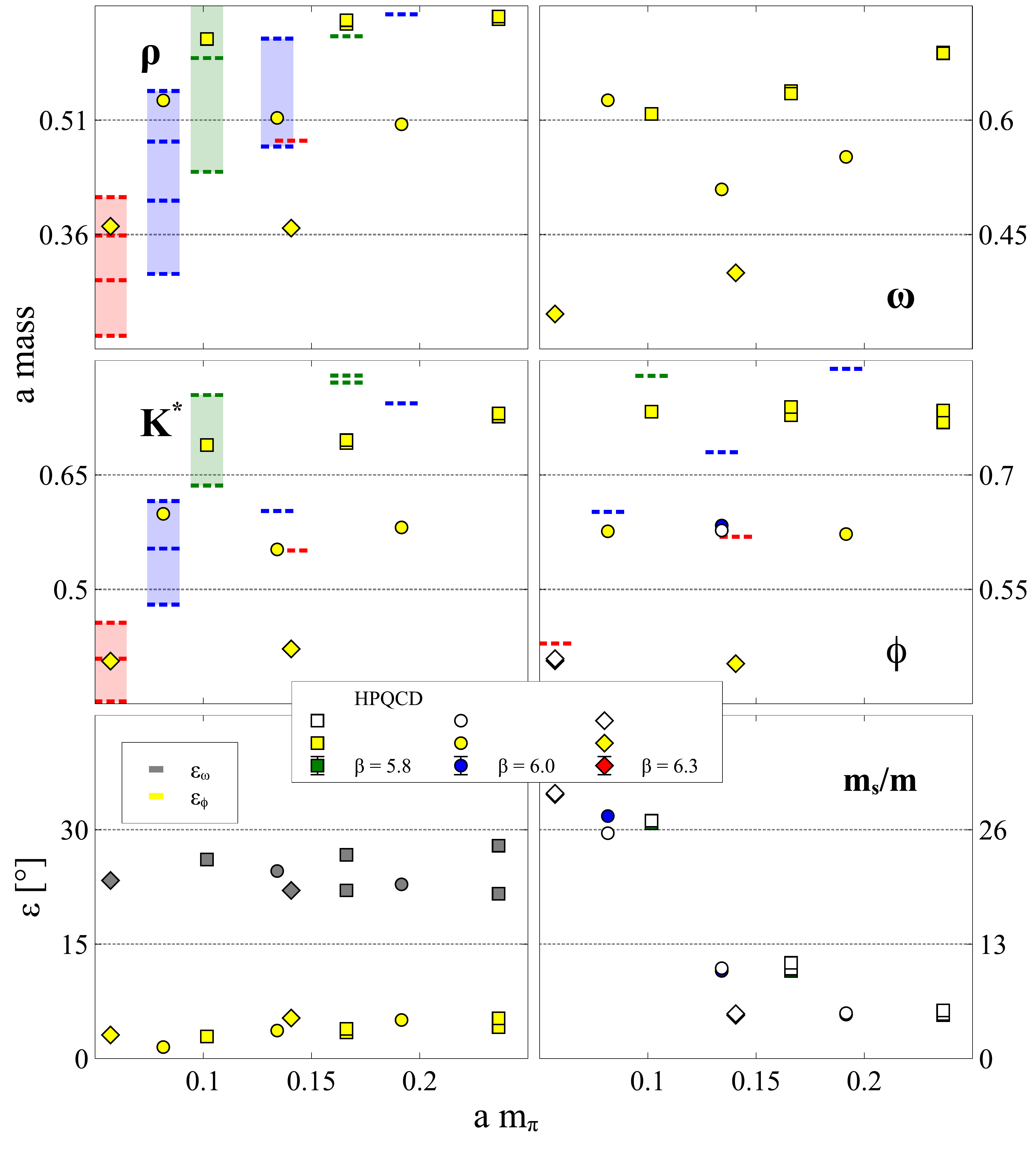}
\caption{Our results from Fit 2 for the vector meson masses, the $\omega-\phi$ mixing angles and quark mass ratios on the HPQCD ensembles. The lattice results are given by green, blue, and  red  filled symbols, where statistical errors are shown only. They are compared to the chiral extrapolation results in open symbols, which are always displayed on top of the lattice symbols. We use yellow or grey colour filled symbols for the cases where there is no corresponding lattice point available yet.  }
\label{fig:4}
\end{figure}

We turn to the decay constants of ETMC \cite{Ottnad:2017bjt}. In Fit 3 the pion decay constants are included in the total chisquare function. The chisquare values in Tab. \ref{tab:parameter:B} are with respect to the latter only. The values are a bit larger than those for HPQCD and CLS. This may hint at somewhat larger discretization effects 
on the ETMC ensembles, in particular on the ensembles with $\beta = 1.95$. 
We note that our lattice scale determination for ETMC in Fit 3 is in the range suggested in \cite{Ottnad:2017bjt}. The chisquare values for Fit 2, which did not consider constraints from ETMC, are based on our lattice scales of Fit. 3.  
In Fig. \ref{fig:2} we show the pion and kaon decay constants in lattice units. The points for the kaon decay constants in the figure are 'pseudo' data obtained from 
\cite{Ottnad:2017bjt} upon a tuning of their wave function factor $Z$. Since only the  leading impact of $Z$ on the kaon decay constants is accessible from  \cite{Ottnad:2017bjt} our points in the figure are subject to additional moderate changes. Our estimates for $Z$ are in the range defined by the values in \cite{Ottnad:2017bjt} based on two distinct methods.  
We find amazing that given our estimates for the wave function we match the quark-mass ratios as provided by ETMC quite accurately. Such ratios are not included in our chisquare function. This is illustrated with Fig. \ref{fig:3}, which shows in addition all vector meson masses and the $\omega-\phi$ mixing angles. Like in our previous work on the PACS-CS, QCDSF-UKQCD and HSC ensembles we foresee a significant quark mass dependence of the $ \omega -\phi$ mixing angles also on the ETMC ensembles. Our predictions for the finite-box vector meson energy levels in Fig. \ref{fig:3} await direct  computations of the latter on the ETMC ensembles. 

We should caution the reader against the case where a given ensemble leads to more than one zero-momentum energy level relevant for a vector meson mass determination. Our self-consistency condition is set up, at this stage, only for a single relevant finite-box energy level. A generalization to more than one level, as it is implied by large volume lattice simulations necessarily, is feasible but beyond the scope of the present work.
In order to delve into this issue we systematically show the first unperturbed two-body scattering levels. From the positions of the latter in Fig. \ref{fig:3} we conclude that our predictions for the energy levels of the vector mesons on the ETMC ensembles are sound.

\begin{figure}[t]
\centering
\includegraphics[width=0.99\textwidth]{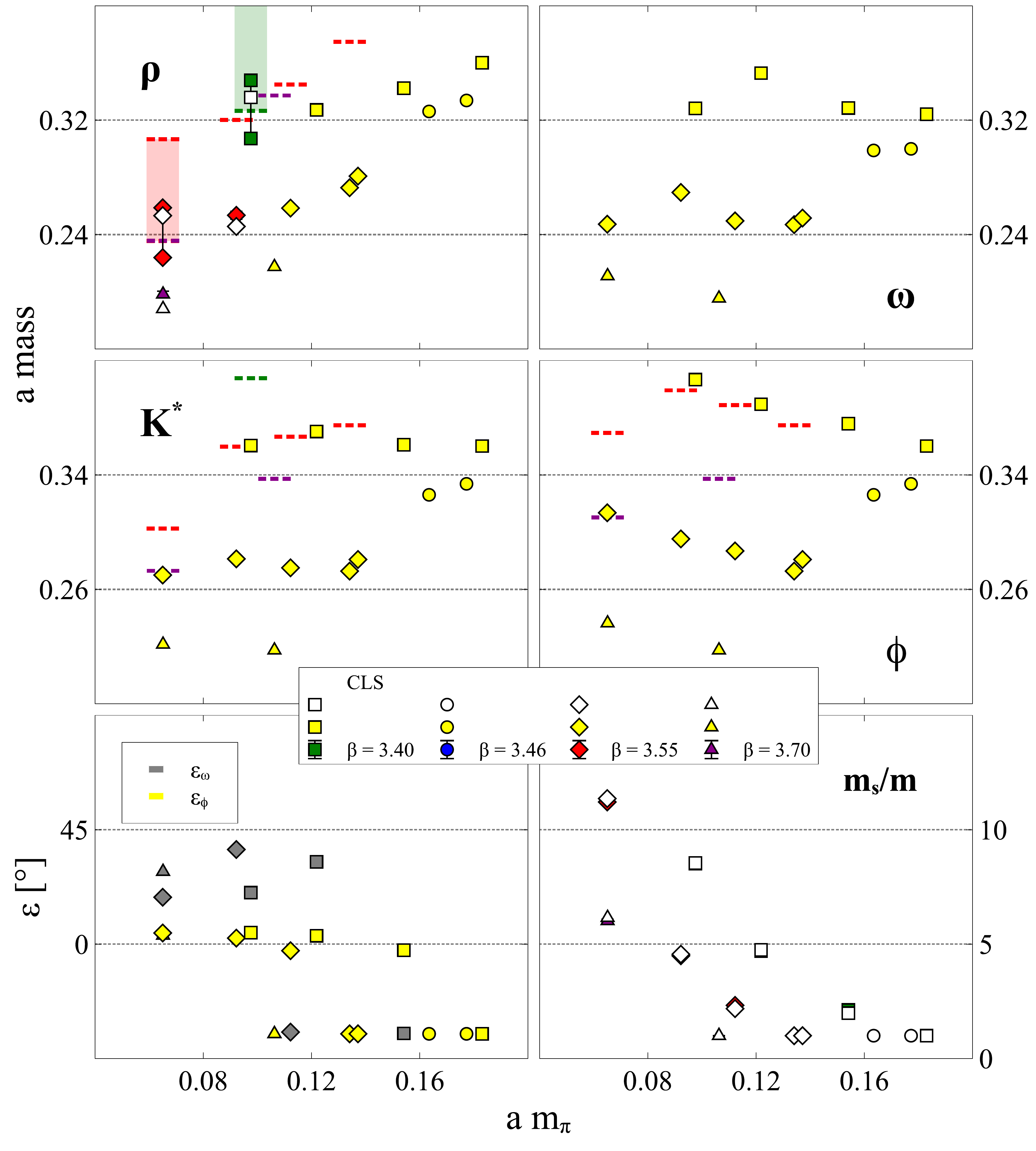}
\caption{Our results from Fit 2 for the vector meson masses, the $\omega-\phi$ mixing angles and quark mass ratios on the CLS ensembles. The lattice results are given by purple, green, blue and red  filled symbols, where statistical errors are shown only. They are compared to the chiral extrapolation results in open symbols, which are always displayed on top of the lattice symbols. We use yellow or grey colour filled symbols for the cases where there is no corresponding lattice point available yet. }
\label{fig:5}
\end{figure}

Our results for the ensembles of HPQCD  are shown in Fig. \ref{fig:4}. Note that for some ensembles a $\phi$ meson energy level is available from HPQCD \cite{Chakraborty:2017hry}. Such levels were considered in Fit 2 and Fit 3 and are accurately reproduced. 
In various cases the scattering levels turn important for the $\rho$ and $K^*$ and therefore our results on such ensembles have to be taken with a grain of salt. 
In Fig. \ref{fig:4} we show the first few unperturbed scattering levels embedded into a shaded area. While we cannot describe the set of expected energy levels in this case, 
our results are still significant. This is so since our self-consistency condition implies an average over the distributed levels. We expect this average to lie close to the most relevant energy level.  
 
We turn to the CLS ensembles as scrutinized in Fig. \ref{fig:5}. The recent results on the $\rho$ meson energy levels in \cite{Andersen:2018mau} were not considered in any of our fits. From the six ensembles analyzed in \cite{Andersen:2018mau} we considered the  four cases C101, D200, N200  and J303. For the two ensembles N200 and J303 there is one significant level for the $\rho$ meson at rest only, and indeed here our 'predictions' are in line with the levels given in \cite{Andersen:2018mau}. For the remaining ensembles C101 and D200 two significant energy levels  are reported on in \cite{Andersen:2018mau}. Those are connected by a solid line in Fig. \ref{fig:5}. Our predicted effective levels are close to the upper one in both cases. Indeed, according to  \cite{Andersen:2018mau} the latter are close to the nominal $\rho$ meson mass. This confirms our expectation that in the presence of more than one relevant energy level we still arrive at significant results. However, to further improve the accuracy of our results it may be useful to generalize  the approach and implement the self-consistency condition for a set of $\rho$ meson energy levels.

In  Fig. \ref{fig:4} and Fig. \ref{fig:5} we show also our predictions for the $\omega-\phi$ mixing angles and the quark mass ratios on the ensembles of HPQCD and CLS. We confirm our previous claim on a striking quark-mass dependence of those mixing angles. 
For both collaborations we recover their quark mass ratios on all considered ensembles quite accurately, despite the fact that none of them was considered in any of our chisquare functions.

\section{Summary}

In this work we considered the pion and kaon decay constants, $f_\pi$ and $f_K$, as evaluated from a chiral SU(3) Lagrangian with dynamical vector meson fields. Our results are based on the one-loop level and the strict isospin limit. It was shown that with Gasser and Leutwyler's $L_4$ and $L_5$ parameters adjusted to the empirical values of $f_\pi$ and $f_K$, corresponding lattice results from HPQCD, CLS and ETMC on ensembles with unphysical quark masses can be reproduced accurately once the effect of dynamical vector meson degrees of freedom are taken into account. At the renormalization point $\mu = 770$ MeV our predicted value for $L_5 = (2 -4)\times10^{-5}$,  is in striking contradiction to conventional estimates based on $\chi$PT studies at the two-loop level. This supports our previous claim that the presence of dynamical vector meson can not be absorbed convincingly into low-orders $\chi$PT approaches.

Our previous result for the chiral SU(3) limit value of the decay constants, which was obtained by global fits to lattice data on the vector meson masses from PACS-CS, QCDSF-UKQCD and HSC, was shown to be consistent with the lattice data on the decay constants from HPQCD, CLS and ETMC.  Our best estimate is $f = (69.1 \pm 1.6)$ MeV.
Quantitative results for the masses of the light vector meson as well as the quark  mass ratios for the ensembles used by  HPQCD and CLS are predicted.

\vskip0.5cm
\centerline{\bf{Acknowledgments}}
\vskip0.5cm
John Bulava and Sinead Ryan are acknowledged for stimulating discussions.  M.F.M. Lutz thanks Kilian Schwarz and Jan Knedlik for support on distributed computing issues.

\bibliographystyle{elsarticle-num}
\bibliography{literature}

\end{document}